\documentclass[onecolumn,showpacs,aps,prd,superscriptaddress,nofootinbib,floatfix]{revtex4}
\usepackage{graphicx}% Include figure files
\usepackage{dcolumn}% Align table columns on decimal point
\usepackage{bm}% bold math

\begin{document}

\title{Update on very light $CP$-odd scalar
in the Two-Higgs-Doublet Model}

\author{F. Larios}
\email{flarios@belinda.mda.cinvestav.mx}
\affiliation{Departamento de F\'{\i}sica Aplicada,
CINVESTAV-M\'erida, A.P. 73,  97310 M\'erida, Yucat\'an, M\'exico}
\author{G. Tavares-Velasco}
\email{gtv@itzel.ifm.umich.mx}
\affiliation{Instituto de F\'{\i}sica y Matem\'aticas, Universidad
Michoacana de San Nicolas de Hidalgo, Apdo. Postal 2-82, C.P.
58040, Morelia, Michoac\'an, M\'exico}
\author{C.-P. Yuan}
\email{yuan@pa.msu.edu}
\affiliation{Department of Physics and Astronomy, Michigan State
University, E. Lansing, MI 48824, USA}
\date{\today}

\begin{abstract}
In a previous work we have shown that a general two-Higgs-doublet
model (THDM) with a very light $CP$-odd scalar  can be compatible
with electroweak precision data, such as the $\rho$ parameter,
BR($b\to s\gamma$), $R_b$, $A_b$, BR($\Upsilon \to A \gamma$),
BR($\eta \to A \gamma$), and $(g-2)$ of muon. Prompted by the
recent significant change in the theoretical status of the latter
observable, we comment on the consequences for this model and
update the allowed parameter region. It is found that the presence
of a very light scalar with a mass of $0.2$ GeV is still
compatible with the new theoretical prediction of the muon
anomalous magnetic moment.
\end{abstract}

%\draft


%%%%%%%%%%%%%%%%%%%%%%%%%%%%%%%%%%%%%%%%%%%%%%%%%%%%%%%%%%%%%%%%%%%%%
%12.60.Fr   Extensions of electroweak Higgs sector
%14.80.Cp   Non-standard-model Higgs bosons
%12.15.Ji   Applications of electroweak models to specific processes
%%%%%%%%%%%%%%%%%%%%%%%%%%%%%%%%%%%%%%%%%%%%%%%%%%%%%%%%%%%%%%%%%%%%%

\maketitle

\section{Introduction}

The possibility of a Higgs boson decaying into a pair of light
$CP$-odd scalars was considered in Ref. \cite{dobrescu}. Although
it is very unlikely that this particle can be accommodated in the
minimal supersymmetric standard model (MSSM), in the light of the
restrictions imposed by the current low-energy data on the
parameters of this model, a very light $CP$-odd scalar $A$ can
still arise in some other extensions of the standard model (SM),
such as the minimal composite Higgs model \cite{composite}, or the
next-to-minimal supersymmetric model \cite{hunter}. Therefore, the
existence of a very light $CP$-odd scalar not only proves new
physics but also casts the most commonly studied MSSM in doubt.
Furthermore, studying the couplings of the light $CP$-odd scalar
to the SM fermions may help discriminating models of electroweak
symmetry breaking -- either a weakly interacting model (e.g., the
next-to-minimal supersymmetric model) or a strongly interacting
model (e.g., the minimal composite Higgs model). Apart from the
above implications arising from the existence of a very light
$CP$-odd scalar, our main interest in studying this particle stems
from the fact that its phenomenology is indeed rather exciting: an
interesting aspect of a light $A$ is that if its mass $M_A$ is
less than twice that of the muon $m_\mu$, i.e. less than about
0.2\,GeV, it can only decay into a pair of electrons ($A\to
e^+e^-$) or photons ($A\to \gamma \gamma$). Hence, the decay
branching ratio BR($A\to \gamma \gamma$) can be sizable.
Consequently, $A$ can behave like a fermiophobic $CP$-odd scalar
and predominantly decay into a photon pair, which would register
in detectors of high energy collider experiments as a single
photon signature when the momentum of $A$ is much larger than its
mass \cite{dobrescu}.

In a previous work  \cite{lighta} we performed an extensive
analysis within the framework of the two Higgs doublet model
(THDM) and found that a very light $CP$-odd scalar can still be
compatible with precision data, such as the $\rho$ parameter,
BR($b\to s\gamma$), $R_b$, $A_b$, BR($\Upsilon \to A \gamma$), and
the muon anomalous magnetic moment $a_\mu$ \cite{lighta}. We
considered different values for $\sin^2(\beta-\alpha)$ and found
the constraints imposed on the remaining parameters of the model,
which we summarize in Table \ref{allconstraints}, where $M_H$
($M_h$) stands for the heavy (light) $CP$-even scalar and
$M_{H^\pm}$ for the charged scalar. As for the soft breaking term
$\mu_{12}$, it is not involved in any of the above observables, so
they cannot be used to constrain it. Since $\mu_{12}$ has no
relevance for the present discussion (the purpose of this work is
to update the bounds derived from the changes in the status of the
theoretical  value of the muon anomaly), we refer the reader to
Ref.~\cite{lighta}, where LEP-2 data were used to set bounds on
this parameter. In obtaining the bounds shown in Table
\ref{allconstraints} we have used the lower value of 110 GeV for
$M_h$. We recall that the LEP-2 direct search bound requires $M_h
> 114.1$\,GeV at the 95\%\,C.L. \cite{EWWG}. However, in the
presence of new physics such a bound can be substantially relaxed.
As explained in Ref. \cite{lighta}, the reason why the LEP-2 bound
($M_h > 114.1$ GeV) does not apply in our model is because this
bound is based on the SM specific value of BR$(h\to b\bar b)$
\cite{lep-2}. In the THDM, the new $h\to AA$ decay mode can
significantly reduce the $h\to b\bar b$ branching ratio. This was
clearly illustrated in the Fig. 9 of Ref.~\cite{lighta} for some
allowed parameter space of the model.   In any case, the new decay
channel ($h\to AA$) registers as a di-photon signature ($h\to
\gamma \gamma$) for which LEP-2 has already set a lower bound. By
taking both the $AA$ and $b \bar b$ decay modes into
consideration, a lower bound for $M_h > 103$ GeV can be
established in our light $A$ scenario \cite{lighta}.

At this point we would like to emphasize that, given the recent
measurements of $a_\mu$  at Brookhaven National Laboratory (BNL)
\cite{brown}, the bounds on new physics effects imposed by the
muon $(g-2)$ data depend largely on the theoretical value
predicted by the SM for the nonperturbative hadronic contribution
to $a_\mu$. In our analysis \cite{lighta}, we followed a
conservative approach and considered various predictions for the
hadronic correction $a^{\rm had}_\mu$
\cite{cly,ay,davier,jegerlehner}, which in fact has been the
source of debate recently \cite{Yndurain,marciano,melnikov}. For
instance, the bounds shown in Table \ref{allconstraints} were
obtained from the calculation presented in Ref. \cite{cly}, which
was the one allowing the largest parameter space.

\begin{table*}
\begin{center}
\caption{Constraints from the low energy data for Type-I and
Type-II THDMs, with $M_A=0.2$ GeV. The old calculation of the
hadronic light by light contribution to $a_{\mu}$, cf. Table
\ref{oldvalues}, was used together with the calculation of Ref.
\cite{cly} for the hadronic vacuum polarization. When
$\sin^2{(\beta-\alpha)}=1$ there is no $M_h$ dependence on $\rho$,
otherwise, we assume $M_h=110$ GeV.}
\begin{tabular}{ccc}
\hline
Constraint & Type-I THDM & Type-II THDM \\
\hline
$(g-2)_\mu$ & $\tan \beta > 0.4 $ & $\tan \beta < 2.6 $ \\
{[$\tan \beta >1$]} $b\to s \gamma$ &
$M_{H^+}>100$ GeV & $M_{H^+}>200$ GeV   \\
{[$0.5< \tan \beta <1$]} $b\to s \gamma$ &
& $M_{H^+}>200\, -350$ GeV  \\
{[$0.6< \tan \beta <1$]} $R_b$ &
$M_{H^+}>200\, -600$ GeV &  $M_{H^+}>200\, -600$ GeV \\
{[$\sin^2 (\beta -\alpha)=1$]} $\Delta \rho$ &
$M_{H}\sim M_{H^+}$ &  $M_H\sim M_{H^+}$ \\
{[$\sin^2 (\beta -\alpha)=0.8$]} $\Delta \rho$ &
$M_{H}\sim 1.2 \, M_{H^+}$ &  $M_H\sim 1.2 \, M_{H^+}$ \\
{[$\sin^2 (\beta -\alpha)=0.5$]} $\Delta \rho$ & $M_{H}\sim 1.7 \,
M_{H^+}$ &  $M_H\sim 1.7 \, M_{H^+}$\\
\hline
\end{tabular}\label{allconstraints}
\end{center}
\end{table*}

\begin{figure}[hbt!]
\includegraphics[width=3in]{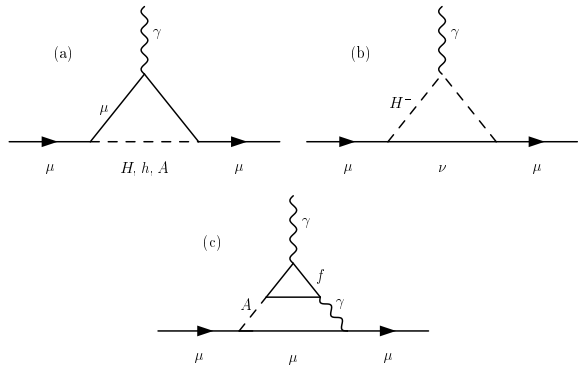}
\caption{\label{magmom}Contribution from the THDM to the anomalous
magnetic moment of muon: (a) neutral Higgs bosons, (b) charged
Higgs boson, and (c) the leading two-loop contribution from the
$CP$-odd scalar.}
\end{figure}

After the completion of our work, it was evident that the latest
precision measurement of $a_\mu$ at BNL \cite{brown} along with
some theoretical predictions for $a^{\rm had}_\mu$ disfavored the
presence of a light $A$ in the THDM. As is well known, the BNL
data opened the prospect for new physics as the experimental value
of $a_\mu$ appeared to be more than $2.6 \,\sigma$ above  the
theory prediction based on some calculations of the hadronic
vacuum polarization. At the one-loop order, a light $CP$-odd
scalar can give a significant negative contribution to $a_\mu$,
making it harder for this type of model to be consistent with
experiment. However, the two-loop calculation can yield a large
correction to the one-loop result as pointed out in \cite{darwin}.
Although this fact seems to contradict perturbation theory, the
unusual situation in which a two-loop diagram can give a
contribution of similar size or even larger than that from the
one-loop diagrams within a perturbative calculation was noted
first by Bjorken and Weinberg when evaluating the Higgs scalar
contribution to the $\mu\to e \gamma$ decay \cite{weinberg}. It is
straightforward to see that this situation also occurs in the
calculation of the Higgs scalar contribution to $a_\mu$. The
reason is that the coupling of the Higgs scalar to the muon enters
twice in the one-loop diagram, whereas at the two-loop level there
appears a diagram in which this coupling enters just once,
together with a line where the Higgs scalar couples to a heavy
fermion pair (see Fig. \ref{magmom}). This gives rise to an
enhancement factor, due to the couplings, that compensates the
suppression factor $g^2/(16\pi^2)$, due to an additional loop. It
turns out that the diagram of Fig. \ref{magmom} (c) gives by far
the most dominant contribution at the two-loop level. Therefore,
we do not expect large uncertainties arising from unknown higher
order terms. In our previous analysis, even when we considered the
two-loop calculation for the $CP$-odd scalar contribution to
$a_\mu$, together with the hadronic correction quoted in Ref.
\cite{brown}, i.e. that by Davier and H\"{o}cker \cite{davier}, we
found that there was no allowed parameter space (in the type-II
THDM) in the $\tan \beta$ vs. $M_A$ plane when $M_A$ was below $3$
GeV. Nevertheless, there were other SM calculations yielding
$a_\mu$ close enough to the experimental value as to allow a very
light $A$.

Since the publication of \cite{brown}, there has been a lot of
controversy regarding the theoretical value of the muon anomalous
magnetic moment. It is evident that before claiming the presence
of any new physics effect, an extensive reexamination of every
contribution to $a_\mu$ is necessary
\cite{Yndurain,marciano,melnikov}. Along these lines, a
reevaluation \cite{knecht} of the hadronic light by light
contribution to $a_\mu$ found a sign error in earlier calculations
\cite{oldlblc} of this contribution, which has resulted in a
significant change of the $a_\mu$ prediction. Once the corrected
value is taken into account, the discrepancy between experiment
and theory reduces down to the level of $1.6 \, \sigma$. In the
light of this result, we believe it is worth revisiting our work
and reexamining our previous bounds.

\section{Allowed parameter range for $M_A$ and $\tan \beta$}

The SM prediction of $a_\mu$ is composed of the following three
parts \cite{Hughes}:

\begin{eqnarray}
a_\mu^{\rm theory} = a_\mu^{\rm QED} \; +\; a_\mu^{\rm
weak} \;+\;  a_\mu^{\rm had},
\end{eqnarray}

\noindent where the electroweak corrections have been computed to
a very good accuracy: they have a combined error of the order of
$5\times 10^{-11}$, which is already about one order of magnitude
smaller than the ultimate goal of the E821 experiment
\cite{brown}.  In contrast, the hadronic contribution $a_\mu^{\rm
had}$ contains the bulk of the theoretical error ($\sim 70\times
10^{-11}$) and can be decomposed into three parts \cite{Hughes},
namely the hadronic vacuum polarization contribution
$a_\mu(\rm{h.v.p.})$, the hadronic light-by-light correction
$a_\mu(\rm{l.b.l})$, and other hadronic higher order terms
$a_\mu(\rm{h.o.})$:

\begin{equation}
a_\mu^{\rm had}=a^{\rm had}_\mu(\rm{h.v.p.})+ a^{\rm
had}_\mu(\rm{l.b.l})+a^{\rm had}_\mu(\rm{h.o.}).
\end{equation}

\begin{table}
\caption{Contributions to the anomalous magnetic moment of muon in
the SM \cite{Hughes}, prior to the discovery of a wrong sign in
the pion pole correction to $a^{\rm had}_\mu({\rm l.b.l})$, which
significantly changed the $a^{\rm theory}_{\mu}$ prediction. All
values are given in units of $10^{-11}$.}
\begin{tabular}{cc}
\hline Contribution&SM prediction \cr\hline $a_\mu^{\rm
QED}$&116584705.7\,(1.8)\cr $a_\mu^{\rm weak}$&151\,(4)\cr $a^{\rm
had}_\mu({\rm l.b.l})$& -79.2\,(15.4) \footnote{This value has
been found to be wrong in Ref. \cite{knecht}.}\cr $a^{\rm
had}_\mu({\rm h.o.})$&-101\,(6)\cr \hline
\end{tabular}\label{oldvalues}
\end{table}

In our previous analysis we used the values shown in Table
\ref{oldvalues} for each contribution to $a_\mu^{\rm theory}$
\cite{Hughes} together with the $a^{\rm had}_\mu(\rm{h.v.p.})$
predictions to be discussed below. In the months following the
publication of our work, a new situation arose: the sign of the
pion pole contribution to the hadronic light by light correction
was found to be wrong \cite{knecht}. Very interestingly, this
contribution alone represents about $70\%$ of the full $a^{\rm
had}_\mu(\rm{l.b.l.})$. It turns out that after correcting this
mistake, the $a^{\rm had}_\mu(\rm{l.b.l.})$ value gets
significantly changed and even its sign gets flipped. As a result,
the discrepancy between the experiment and theory reduces down to
the level of $1.6 \, \sigma$. Subsequent publications have
confirmed this finding \cite{kinoshita,bijnens,blokland}. In Table
\ref{L.B.L.amu} we list the most recent evaluations of $a^{\rm had
}_\mu(\rm{l.b.l.})$. In addition, there is one more calculation
that is based on chiral perturbation theory \cite{wise}:

\begin{eqnarray}
a^{\rm had}_\mu({\rm l.b.l.}) = \left(\; 55^{+50}_{-60} \;
+\; 31 \hat C \;\right) \; \times 10^{-11}, \nonumber
\end{eqnarray}

\noindent where $\hat C$ is an unknown low-energy constant that
parametrizes some subdominant terms. We will not consider this
result here but only mention it as an example of a calculation
that is still open to debate. For the purpose of this work we will
take an average of the top three results shown in Table
\ref{L.B.L.amu} and study the consequences on the allowed
parameter space of the THDM.

\begin{table}
\caption{The most recent evaluations of the hadronic light by
light contribution to $a^{\rm had}_\mu({\rm l.b.l.})$. These
corrected values contrast with the wrong one shown in Table
\ref{oldvalues}.}
\begin{tabular}{cc}
\hline Author&$a^{\rm had}_\mu({\rm l.b.l.})\times 10^{11}$
\cr\hline KN \cite{knecht}&83\,(12)\cr
HK\cite{kinoshita}&89\,(15)\cr BPP\cite{bijnens}& 83\,(32)\cr BCM
\cite{blokland}&56\footnote{This value accounts only for the pion
pole contribution.}\cr \hline
\end{tabular}
\label{L.B.L.amu}
\end{table}

After introducing the corrected value of $a^{\rm had}_\mu({\rm
l.b.l.})$, the sum of all the contributions to $a_\mu^{\rm
theory}$ except $a_\mu^{\rm had}(\rm{h.v.p.})$ is:

\begin{equation}
a_\mu^{\rm theory}- a_\mu^{\rm had}(\rm{h.v.p.})= 116584845.3\, (
17.1) \times 10^{-11}, \label{amuehvo}
\end{equation}

\noindent where the errors have been composed quadratically.
\footnote{Throughout this work we will systematically compose the
errors in quadrature.} As for the $a_\mu({\rm h.v.p})$ term, its
evaluation has also been the source of renewed interest lately.
\footnote{For a summary of the most recent evaluations of $a^{\rm
had}_\mu$, see Refs. \cite{troconiz} and \cite{narison}.} As in
our previous work, here we will use a conservative approach and
consider some representative evaluations of $a_\mu({\rm h.v.p})$.
In the second column of Table \ref{h.v.p.amu} we show some of the
most recent results, which were compiled in \cite{troconiz},
whereas in the third column we show the full theory prediction,
which is obtained after adding up each value in the second column
to Eq. (\ref{amuehvo}).

\begin{table*}
\caption{Some of the most recent calculations of $a^{\rm
had}_\mu({\rm h.v.p.})$ together with the respective theory
prediction $a_\mu^{\rm theory}$ and the discrepancy $\Delta a_\mu$
between experiment and theory. The last column represents the
bounds on any new physics contribution $a_\mu^{\rm NP}$ at the
95\% C.L. All of the values are given in units of $10^{-11}$.}
\begin{tabular}{ccccc}
\hline Author&$a^{\rm had}_\mu({\rm h.v.p.})$&$a_\mu^{\rm
theory}$&$\Delta a_\mu$& Allowed range for $a_\mu^{\rm
NP}$\cr\hline ADH
\cite{alemany}&7011(94)&116591856.3\,(95.54)&166.7\,(179.53)&[-185\,--\,519]\cr
DH
\cite{davier}&6924(62)&116591769.3\,(64.31)&253.7\,(164.92)&[-70\,--\,577]\cr
J
\cite{jegerlehner}&6974(105)&116591833.3\,(112.3)&189.7\,(188.8)&[-180\,--\,560]\cr
N
\cite{narison}&7031(77)&116591876.3\,(78.88)&146.7\,(171.25)&[-189\,--\,482]\cr
TY
\cite{troconiz}&6952(64)&116591797.3\,(66.25)&225.7\,(165.81)&[-99\,--\,551]\cr
\hline
\end{tabular}
\label{h.v.p.amu}
\end{table*}

As for the experimental value $a_\mu^{\rm exp}$, the data obtained
during the 1999 running period combined with previous measurements
give \cite{brown}

\begin{equation}
a_\mu^{\rm{exp}}=116592023\,(152)\times 10^{-11}. \label{amuexp}
\end{equation}

\noindent One thus can obtain the discrepancy between experiment
and theory $\Delta a_\mu=a_\mu^{\rm exp} -a_\mu^ {\rm theory}$ for
each different evaluation of  $a_\mu({\rm h.v.p})$, as shown in
the fourth column of Table \ref{h.v.p.amu}. Finally, if we assume
that the discrepancy between theory and experiment is to be
ascribed to new physics effects, we can obtain the bounds shown in
the last column of the same Table for the new physics contribution
to the anomalous magnetic moment at the 95 \% C.L., which is
denoted by $a_\mu^{\rm NP}$. Those bounds on $a_\mu^{\rm NP}$
should be compared to those used in our previous analysis, cf. Eq.
(2) in Ref. \cite{lighta}.

\begin{figure}[hbt!]
\includegraphics[width=3in]{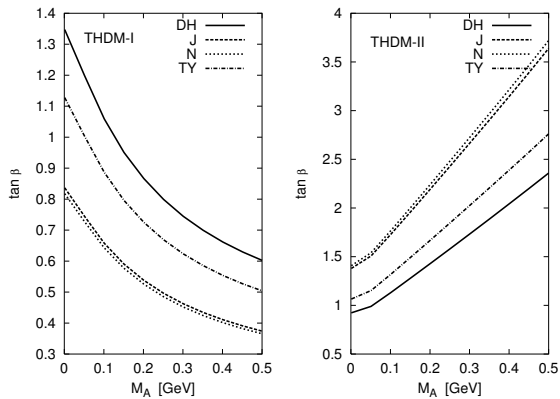}
\caption{\label{gm2tbma}The regions (above the curves for type-I
and below the curves for type-II THDM) in the $\tan\beta$ versus
$M_A$ plane allowed by the $a_\mu$ data at the 95\% CL. Four
different curves are displayed depending whether the SM prediction
is obtained from the DH, J, N or TY calculation of $a^{\rm
had}_\mu({\rm h.v.p.})$. The two-loop contribution from the light
$A$ has been used.}

\end{figure}

\begin{figure}[hbt!]
\includegraphics[width=3in]{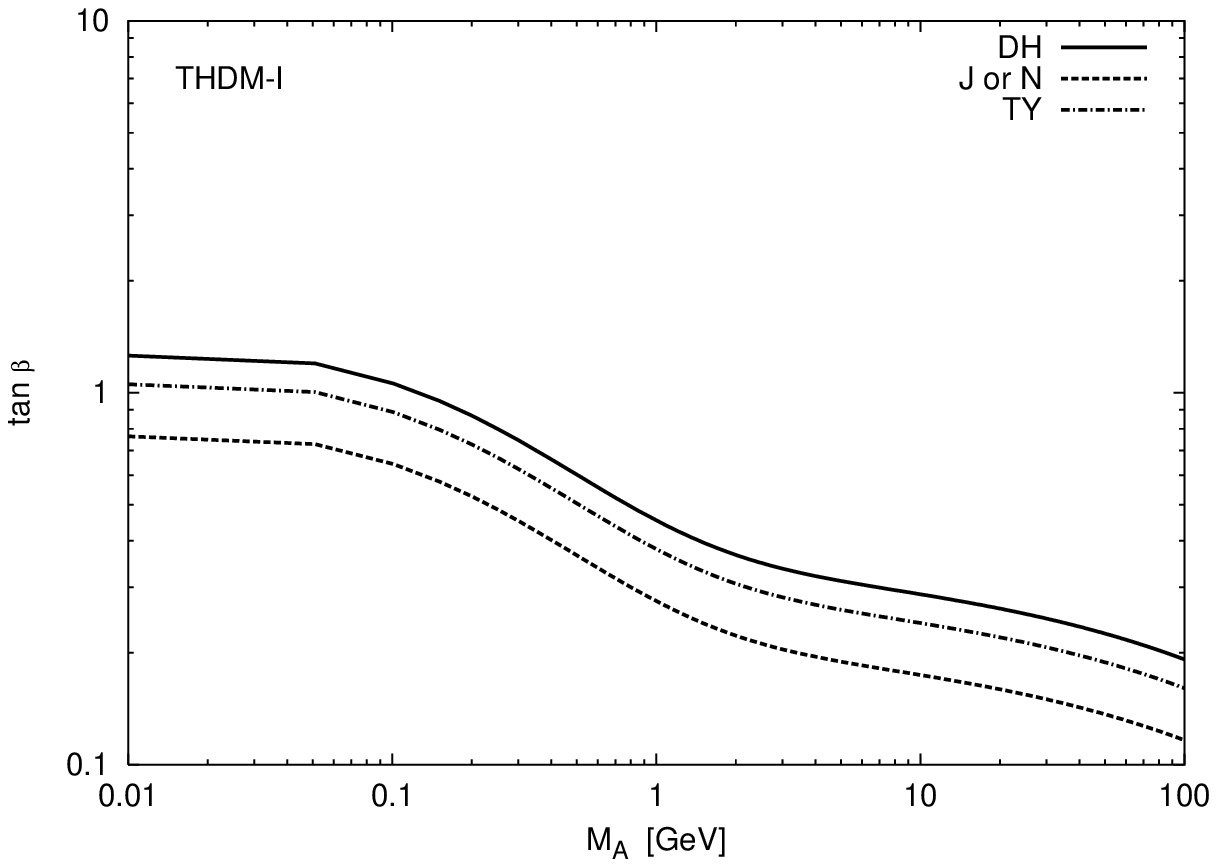}
\caption{\label{gm2tbma-1}The region (above the curves)
 in the $\tan\beta$ versus $M_A$ plane of a
type-I THDM allowed by the $a_\mu$ data at the 95\% CL. The
allowed regions based on the DH, J, N and TY calculations are
above the curves. The two-loop contribution from the light $A$ has
been used.}
\end{figure}

\begin{figure}[hbt!]
\includegraphics[width=3in]{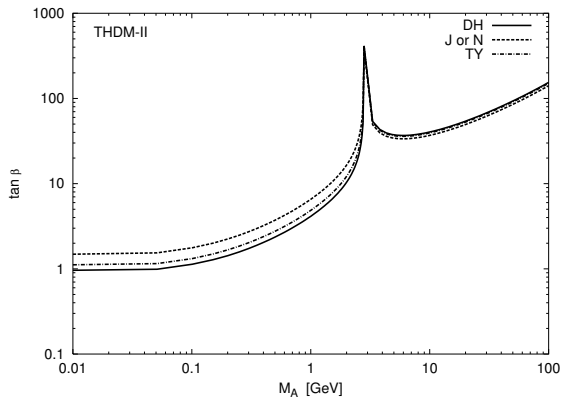}
\caption{\label{gm2tbma-2}The regions (below the curves) in the
$\tan\beta$ versus $M_A$ plane of a type-II THDM allowed by the
$a_\mu$ data at the 95\% CL. The allowed regions based on the DH,
J, N and TY calculations are below the curves.  The two-loop
contribution from the light $A$ has been used.}
\end{figure}

\begin{figure}[hbt!]
\includegraphics[width=3in]{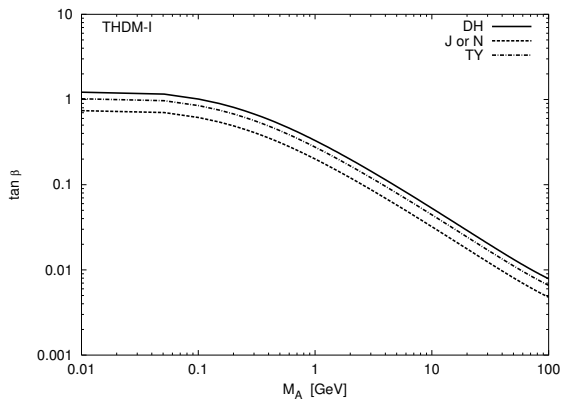}
\caption{\label{gm2tbma-one-1}Same as Fig. \ref{gm2tbma-1}, but
only the one-loop contribution from the light $A$ is considered.}

\end{figure}

\begin{figure}[hbt!]
\includegraphics[width=3in]{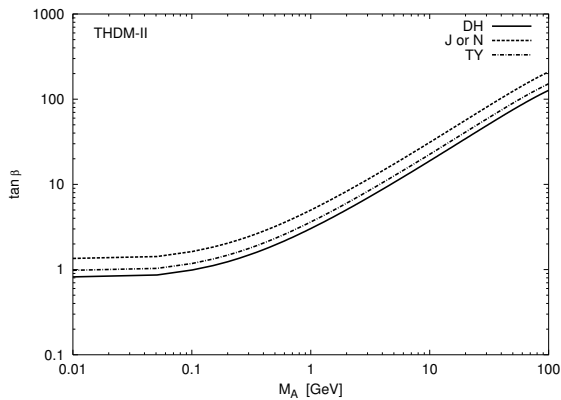}
\caption{\label{gm2tbma-one-2}Same as Fig. \ref{gm2tbma-2}, but
only the one-loop contribution from the light $A$ is considered.}

\end{figure}

Given the new bounds on $a_\mu^{\rm NP}$, we update the constraint
imposed by it on the $\tan\beta-M_A$ plane within the THDM. The
analytical expressions for the contribution of either a $CP$-even
or a $CP$-odd scalar (Fig. \ref{magmom}) can be found in Appendix
A of Ref. \cite{lighta}. We will use the two-loop calculation for
the contribution from the $CP$-odd scalar \cite{darwin}. In order
to satisfy the bounds shown in Table \ref{allconstraints}, we are
assuming that the remaining four Higgs scalars are much heavier
than the $CP$-odd scalar $A$, so their contribution to $a_\mu$
turns out to be negligibly small as compared to that coming from
the latter. Also, we are considering that $\sin^2 (\beta-\alpha)=
1$. The reason why we make this choice is because in our scenario
with a very light $CP$-odd scalar the most convenient way to meet
the constraint imposed by the $\rho$ parameter is to have $M_H$
and $M_{H^+}$ nearly degenerate and $\sin^2 (\beta-\alpha)$ close
to 1 \cite{lighta}. For comparison purposes, we will analyze the
bounds arisen from the theoretical predictions based on the DH
\cite{davier}, J \cite{jegerlehner}, N \cite{narison} and TY
\cite{troconiz} calculations of $a^{\rm had}_\mu({\rm h.v.p.})$,
which are the most representative and recent ones. We would like
to note that, as observed through Fig. \ref{gm2tbma} to Fig.
\ref{gm2tbma-one-2}, the bounds from the J \cite{jegerlehner} and
N \cite{narison} calculations are almost indistinguishable.

In Figs. \ref{gm2tbma}-\ref{gm2tbma-2} we show the allowed regions
in the $\tan\beta-M_A$ plane for both types of THDMs.  In Fig.
\ref{gm2tbma}, which shows the low $M_A$ regime, it can be seen
clearly that even if one considers the DH calculation of $a^{\rm
had}_\mu$, which is the one with the smallest error, there is
still the possibility of having a $CP$-odd scalar with a mass of
the order of $0.2$ GeV in the type-II THDM as long as
$\tan\beta<1.43$, whereas for a type-I THDM $\tan \beta$ has to be
greater than $0.87$. This is a very significant change with
respect to the results obtained when using the old (uncorrected)
value of $a^{\rm theory}_\mu$. In that case, the DH calculation
did not allow for a light $CP$-odd scalar in either type of
THDM, though other calculations did allow such a possibility.

As stated above, so far our results have been derived from the
two-loop contribution from the $CP$-odd scalar to $a^{\rm
NP}_\mu$. It is also interesting to repeat the above analysis
using only the one-loop calculation for $a^{\rm NP}_\mu$. Its
result is depicted in Figs. \ref{gm2tbma-one-1} and
\ref{gm2tbma-one-2}. The old (uncorrected) theory prediction based
on the DH calculation required any new physics contribution to
$a_\mu$ to be positive. However, the one-loop contribution from a
light $CP$-odd scalar is always negative. Therefore, the old SM
theory prediction for $a_\mu$ combined with the THDM one-loop
correction strongly disfavored the existence of a very light
$CP$-odd scalar. This is to be contrasted with the conclusion
drawn from the corrected value of $a^{\rm theory}_\mu$. In that
case, there is indeed an allowed region of $\tan\beta$ when
$M_A\sim 0.2$ GeV, though this region is smaller than  the one
allowed by the two-loop calculation of $a^{\rm NP}_\mu$ (cf. Figs.
3 and 5, and Figs. 4 and 6). As shown in Fig. \ref{gm2tbma-2},
there is an interesting feature in the $\tan\beta$ versus $M_A$
plane of a type-II THDM when $M_A$ is around 2.6\,GeV. It is
because for $M_A \sim 2.6$\,GeV, the two-loop contribution from a
light $CP$-odd scalar becomes as large as the respective one-loop
contribution but with an opposite sign, so the total effect
cancels.

\subsection{Bounds on $\tan \beta$ from meson decays}

For completeness we now turn to analyze the bounds obtained on
THDMs with a very light $CP$-odd scalar from meson decays. A very
light Higgs scalar ($CP$-odd or $CP$-even) can be a decay product
of some hadrons, like the $\eta$ and $\Upsilon$ mesons. For the
latter, a measured upper bound to the $X+\gamma$ decay channel has
been set \cite{pdg} that can be used to constrain the $A\,\bar b\,
b$ coupling. Denote the Yukawa coupling of $A\,\bar b\, b$ to be
$k_d m_b/v$, with $k_d =\tan \beta$ ($\cot\beta$) in the type-II
(type-I) model. Then, the data of the meson decay $\Upsilon \to
\gamma + X$ requires $k_d < 1$. (We refer the reader to Refs.
\cite{lighta, hunter} for a detailed discussion.)

As shown in Ref. \cite{prades}, there is another decay process
that can strongly constrain $\tan \beta$, namely  $\eta \to \pi
S$, where $S$ is a very light $CP$-even scalar. Those results can
be translated into the case of a $CP$-odd scalar. In particular,
the experimental upper limit

\begin{equation}
{\mathrm{BR}} (\eta \to \pi^0 e^+ e^-) \leq 5\times 10^{-5}
\end{equation}

\noindent can be used to obtain the following constraint on a THDM
$CP$-odd scalar with mass $M_A$ lying in the range $2m_e\leq M_A
\leq 2 m_\mu$:

\begin{equation}
(k_d - k_u)^2
\lambda\left(1,\frac{m^2_\pi}{m^2_\eta},\frac{m^2_A}{m^2_\eta}
\right) \leq 1.5
\end{equation}

\noindent where $k_u$ is $\cot \beta$ for either type-I or type-II
THDM and $k_d$ has been defined above. The function $\lambda$ is
given by $\lambda^2 (a,b,c) = a^2 + b^2 + c^2 - 2ab - 2 ac - 2
bc$. From here we can conclude that $\cot \beta \geq 0.65$ for
type-I THDM and $0.55 \leq \tan \beta \leq 1.8$ for type-II THDM.
We thus can confirm that the hadron decay data together with the
muon $(g-2)$ measurement require $\tan \beta$ to be of order 1 if
there exists a very light pseudoscalar with a mass smaller than $2
m_\mu$.

\section{Overall description of the general THDM with
a light $A$}

Once the allowed parameter range for $\tan \beta$ and $M_A$ has
been updated, there remains five other parameters to consider: the
$CP$-even neutral Higgs mixing angle $\alpha$, the soft breaking
term $\mu_{12}$ and the three other Higgs masses: $M_h$, $M_H$ and
$M_{H^+}$. Since we already know that $\tan \beta$ has to be of
order 1 we can address the status of the charged Higgs mass
$M_{H^+}$ independently of the other parameters. It turns out that
both the $b\to s \gamma$ and the $R_b$ data require $H^+$ to be
considerably heavy \cite{lighta, gambino}:

\begin{equation}
M_{H^+} \gtrsim 350\; {\rm GeV} \; .
\end{equation}

Such a high lower bound for the $H^+$ mass affects the allowed
values of the mixing angle $\alpha$. In Ref. \cite{lighta} we show
that the $\rho$ parameter requires $M_H$ and $M_{H^+}$ to be very
correlated depending on the value of $\sin^2 (\beta - \alpha)$. In
fact, if a very light $CP$-odd scalar is to be allowed, the
easiest way to satisfy the bound imposed by $\rho\sim 1$ is to
have $M_H$ and $M_{H^+}$ degenerate and $\sin^2(\beta-\alpha)=1$.
With this choice, $M_h$ is not restricted since it does not
contribute to the $\rho$ parameter. As we consider values of
$\sin^2 (\beta -\alpha)$ smaller than 1, it turns out that $\rho$
is very sensitive to the masses of $H$ and $H^+$. For instance, if
$\sin^2 (\beta -\alpha) = 0.5$, $M_H$ must be at least of the
order of 500 GeV \cite{lighta}. Generally speaking, our conclusion
on the bounds on a very light $CP$-odd scalar in the THDM does not
change significantly for $0.5< \sin^2 (\beta -\alpha) < 1$ as long
as the other Higgs bosons in the model are heavy enough. For a
very small value of $\sin^2 (\beta -\alpha)$, much less than 0.5,
the $\rho$-parameter data would have required the mass of $H$ to
be at the TeV order.

\section{Conclusion}

In conclusion, with the recent correction to the SM prediction of
$a_\mu$, the current muon $(g-2)$ data, together with other
precision data (cf. Table I), still allows a light ($M_A\sim
0.2$\,GeV) $CP$-odd scalar boson in the THDM. Due to this new
development in the SM theory calculation of muon's $(g-2)$ , the
allowed range of $\tan\beta$ in the Type-I or Type-II THDM is
modified, and our result is summarized in Table \ref{gm2tb_all}.
It is interesting to note that the phenomenology at high energy
colliders predicted by the THDM with a light $CP$-odd Higgs boson
is dramatically different from that predicted by the usual THDM in
which the mass of the $CP$-odd scalar is at the weak scale. A
detailed discussion on this point can be found in Ref.
\cite{lighta}. In particular, various potential discovery modes
were studied in there: it was found that the Fermilab Tevatron,
the CERN large hadron collider (LHC) and the planned $e^+e^-$
linear collider (LC) have a great potential to either detect or
exclude a very light $A$ in the THDM.

\begin{table}
\begin{center}
\caption{Constraints on $\tan\beta$ from the muon $(g-2)$ data for
Type-I and Type-II THDM, with $M_A=0.2$ GeV, based on various SM
theory predictions of $a^{\rm had}_\mu({\rm h.v.p})$. The two-loop
contribution for the $CP$-odd scalar has been used.}
\begin{tabular}{ccc}
\hline
Theory prediction & Type-I THDM & Type-II THDM \\
\hline
DH \cite{davier} & $\tan \beta > 0.87 $ & $\tan \beta < 1.43$ \\
J \cite{jegerlehner}  & $\tan \beta > 0.54 $ & $\tan \beta < 2.19$ \\
N \cite{narison}  & $\tan \beta > 0.53 $ & $\tan \beta < 2.24$\\
TY \cite{troconiz} & $\tan \beta > 0.73 $ & $\tan \beta < 1.67$\\
\hline
\end{tabular}\label{gm2tb_all}
\end{center}
\end{table}

Finally, we note that while a light $CP$-odd scalar in THDM is
still compatible with all the precision data, it has been shown
recently in Ref. \cite{Hagiwara} that a light $CP$-odd scalar in
the MSSM will violate the constraint derived from the $Z b {\bar
b}$ coupling. This is because in the MSSM, the masses of the five
Higgs bosons are related by the mass relations required by
supersymmetry. Hence, with a light $CP$-odd scalar, the mass of
the other Higgs bosons cannot be arbitrary large, and it is
difficult to yield the decoupling limit when calculating low
energy observables.

\vskip0.5in \leftline{\bf\large{Note added:}} \vskip0.2in

\begin{figure}[hbt!]
\includegraphics[width=3in]{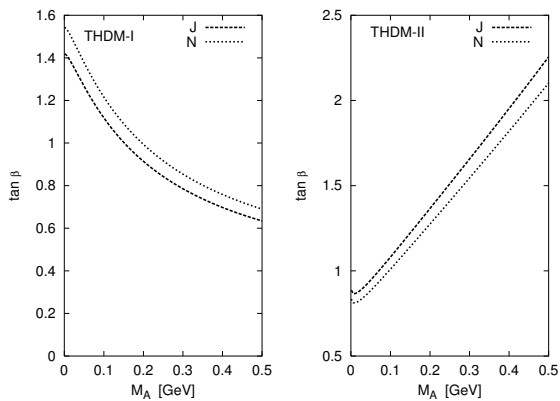}
\caption{\label{newgm2tbma}Same as Fig. \ref{gm2tbma}, but with
the latest experimental data from the muon $(g-2)$ collaboration
\cite{newBNL}. There is no allowed region in this range of
parameters according to the DH \cite{davier} and TY
\cite{troconiz} calculations.}
\end{figure}

\begin{figure}[hbt!]
\includegraphics[width=3in]{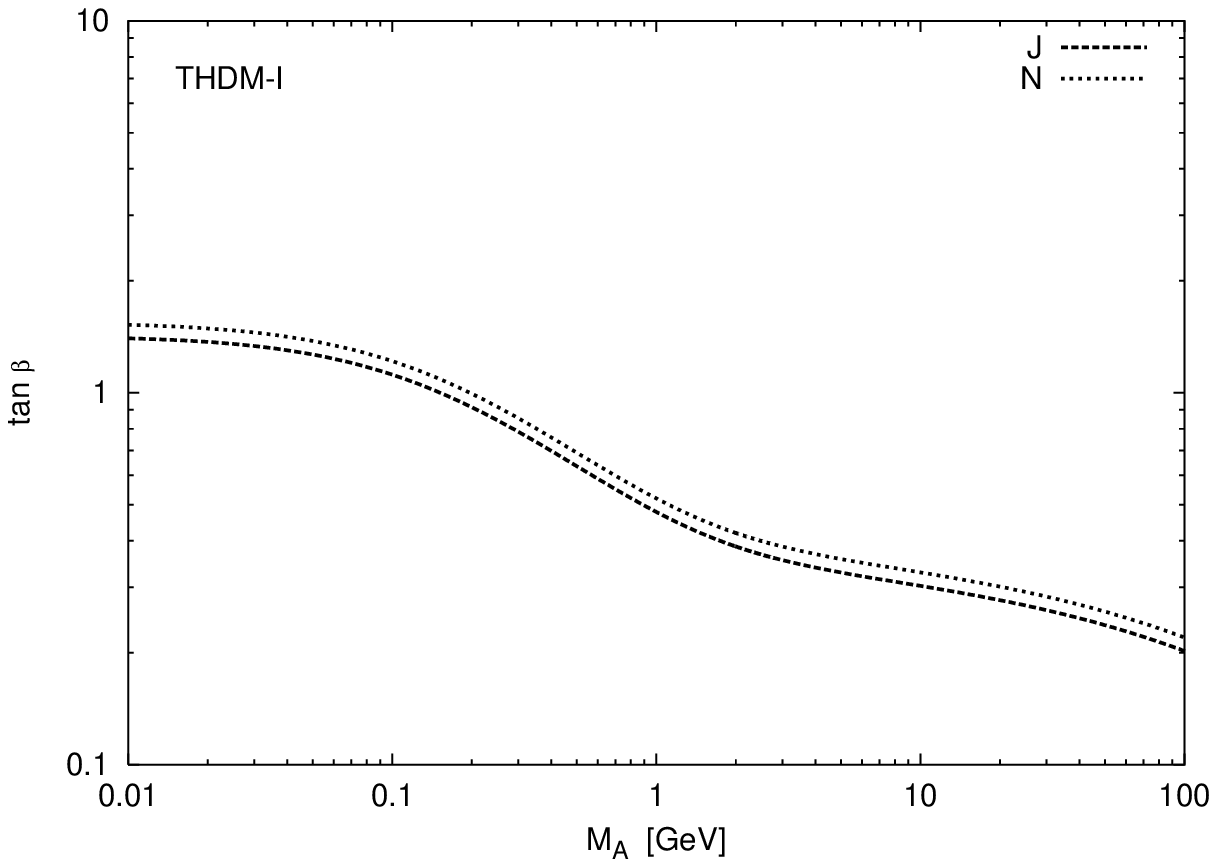}
\caption{\label{newgm2tbma-1}Same as Fig. \ref{gm2tbma-1}, but
with the latest experimental data from the muon $(g-2)$
collaboration \cite{newBNL}.  There is no allowed region in this
range of parameters according to the DH \cite{davier} and TY
\cite{troconiz} calculations.}
\end{figure}

\begin{figure}[hbt!]
\includegraphics[width=3in]{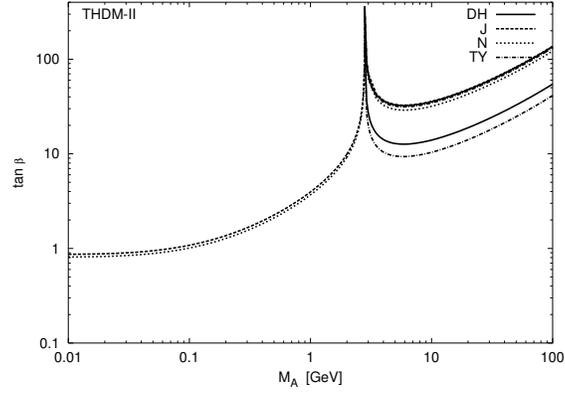}
\caption{\label{newgm2tbma-2}Same as Fig. \ref{gm2tbma-2}, but
with the latest experimental data from the muon $(g-2)$
collaboration  \cite{newBNL}.  The region allowed by the DH
\cite{davier} and TY \cite{troconiz} calculations is bounded by
the respective lines.}
\end{figure}

\begin{figure}[hbt!]
\includegraphics[width=3in]{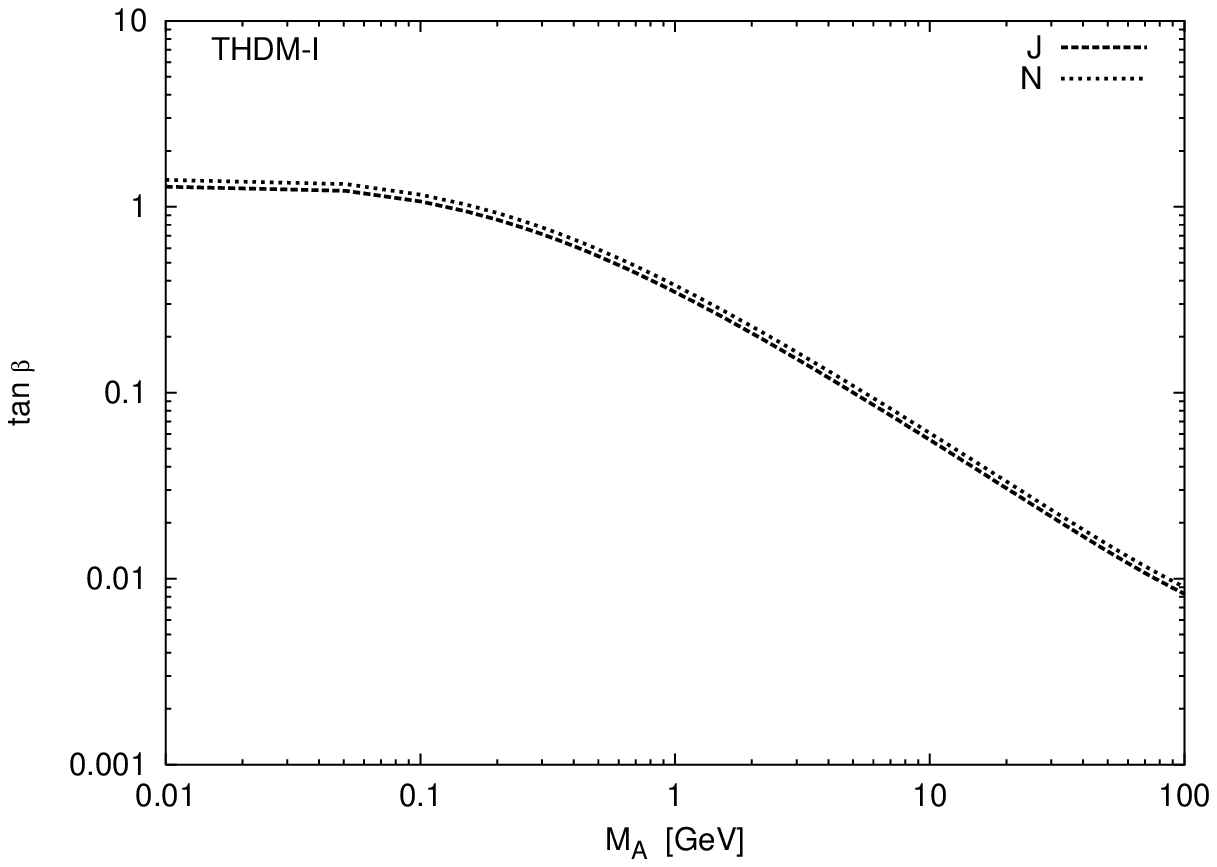}
\caption{\label{newgm2tbma-one-1}Same as Fig. \ref{gm2tbma-one-1},
but with the latest experimental data from the muon $(g-2)$
collaboration \cite{newBNL}.  There is no allowed region in this
range of parameters according to the DH \cite{davier} and TY
\cite{troconiz} calculations.}
\end{figure}

\begin{figure}[hbt!]
\includegraphics[width=3in]{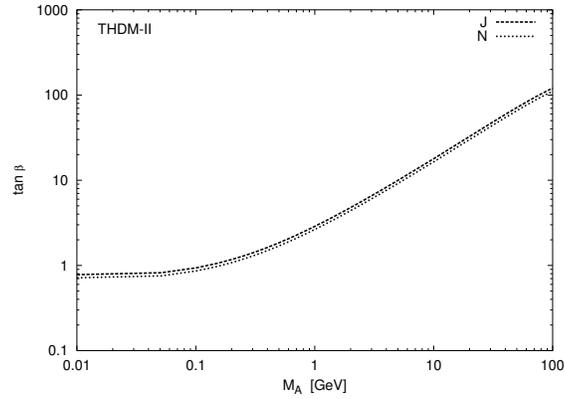}
\caption{\label{newgm2tbma-one-2}Same as Fig. \ref{gm2tbma-one-2},
but with the latest experimental data from the muon $(g-2)$
collaboration considered \cite{newBNL}.  There is no allowed
region in this range of parameters according to the DH
\cite{davier} and TY \cite{troconiz} calculations.}
\end{figure}

During the review process of this manuscript, the muon $(g-2)$
collaboration announced a new result based on data collected in
the year 2000 \cite{newBNL}, in which the experimental uncertainty
has been reduced to one half that of the previous measurement
while the central value of $a_\mu^{\rm{exp}}$ remains about the
same. [The new data yields $a_\mu^{\rm{exp}}=11 659 204(7)(5)
\times 10^{-10}\quad(0.7\; {\mathrm{ppm}})$.] According to the
latest experimental data, we have updated Figs. \ref{gm2tbma} to
\ref{gm2tbma-one-2} in this paper to Figs. \ref{newgm2tbma} to
\ref{newgm2tbma-one-2}. The new data suggests that a very light
$CP$-odd scalar is not allowed in the type-I or type-II THDM based
on the SM calculation done by DH \cite{davier} and TY
\cite{troconiz}. However, based on the N \cite{narison} and J
\cite{jegerlehner} calculations, a very light $CP$-odd scalar is
still possible though the allowed parameter space of the THDM has
been tightly constrained.

\vspace{1cm}
%%%%%%%%%%%%%%%%%%%%%%%%%%%%%%%%%%%%%%%%%%%%%%%%%%
\begin{acknowledgments}
FL would like to thank Conacyt and SNI (M\'exico) for support. GTV
acknowledges support from SEP-PROMEP. The work of CPY was
supported in part by NSF grant PHY-0100677.
\end{acknowledgments}

%%%%%%%%%%%%%%%%%%%%%%%%%%%%%%%%%%%%%%%%%%%%%%%%%

\end{document}